\documentclass[12pt,english,aps,prl, showpacs,preprintnumbers,amsmath,amssymb,showkeys]{revtex4-2}

\usepackage[T1]{fontenc}
\usepackage[latin9]{inputenc}
\usepackage{color}
\usepackage{units}
\usepackage{amssymb}
\usepackage{graphicx}
\usepackage{esint}
\usepackage{bm}
\usepackage{natbib}
\usepackage{amsmath} 
\usepackage{ulem}
\usepackage{hypernat}
\usepackage{xcolor}
\usepackage[colorlinks=true, citecolor=blue]{hyperref}
\usepackage{chemformula} % Formula subscripts using \ch{}

\begin{document}
	
	\title{Experimental observation of spin--split energy dispersion in high-mobility single-layer graphene/\ch{WSe2} heterostructures}
	
	\author{Priya Tiwari$^1$, Mohit Kumar Jat$^1$, Adithi Udupa$^2$, Deepa S. Narang$^1$,  Kenji Watanabe$^3$, Takashi Taniguchi$^4$, Diptiman Sen$^{1,2}$, Aveek Bid$^{1*}$}
	
	\address{$^1$Department of Physics, Indian Institute of Science, Bangalore 560012, India \\
		$^2$Centre for High Energy Physics, Indian Institute of Science, Bangalore 560012, India \\
		$^3$ Research Center for Functional Materials, National Institute for Materials Science, 1-1 Namiki, Tsukuba 305-0044, Japan \\
		$^4$ International Center for Materials Nanoarchitectonics, National Institute for Materials Science, 1-1 Namiki, Tsukuba 305-0044, Japan\\
		$^*$ aveek@iisc.ac.in\\}

	\begin{abstract}
Proximity-induced spin-orbit coupling in graphene has led to the observation of intriguing phenomena like time-reversal invariant $\mathbb{Z}_2$ topological phase and spin-orbital filtering effects. An understanding of the effect of spin-orbit coupling on the band structure of graphene is essential if these exciting observations are to be transformed into real-world applications. In this research article, we report the experimental determination of the band structure of single-layer graphene (SLG) in the presence of strong proximity-induced spin-orbit coupling. We achieve this in high-mobility hBN-encapsulated SLG/\ch{WSe2} heterostructures through measurements of quantum oscillations. We observe clear spin-splitting of the graphene bands along with a substantial increase in the Fermi velocity. Using a theoretical model with realistic parameters to fit our experimental data, we uncover evidence of a band gap opening and band inversion in the SLG. Further, we establish that the deviation of the low-energy band structure from pristine SLG is determined primarily by the valley-Zeeman SOC and Rashba SOC, with the Kane-Mele SOC being inconsequential. Despite robust theoretical predictions and observations of band-splitting, a quantitative measure of the spin-splitting of the valence and the conduction bands and the consequent low-energy dispersion relation in SLG was missing -- our combined experimental and theoretical study fills this lacuna.
	\end{abstract}
	
	% Uncomment for keywords
\vspace{2pc}
	\maketitle

	\section{Introduction}
	
	Graphene has attracted much attention due to a plethora of remarkable electronic properties like Dirac energy dispersion, relativistic effects, half-integer quantum Hall effect, and Klein tunneling. Additionally, its van der Waals heterostructures with other 2-dimensional materials~\cite{Geim2013,doi:10.1146/annurev-matsci-070214-020934,C4NR03435J,doi:10.1002/adfm.201706587} host several single-particle and emergent correlated states that are topologically non-trivial~\cite{tiwari2020observation, cao2018unconventional, hatsuda2018evidence, cao2018correlated}. The ability to precisely transfer and align these atomically thin planar structures into high-quality heterostructures promises outstanding opportunities for both fundamental and applied research~\cite{doi:10.1063/1.4919793, tiwari2021electric, avsar2014spin}. This has made theoretical and experimental studies of several aspects of graphene-based van der Waals heterostructures of great contemporary interest~\cite{ingla2021electrical, sierra2021van, zollner2020swapping, zollner2020scattering, cao2018correlated, safeer2019room, avsar2014spin, tiwari2021electric, cysne2018quantum, C7CS00864C, wang2015strong}.
	
	With a long spin-relaxation length of several $\mu$m at room temperature, graphene appears  a perfect base for low-power spintronics devices~\cite{RevModPhys.92.021003,benitez2020tunable}. However, its extremely weak intrinsic spin-orbit coupling (SOC) strength makes spin manipulation challenging. Decorating the surface of graphene with heavy atoms (such as topological nanoparticles)~\cite{hatsuda2018evidence, weeks2011engineering} or weak hydrogenation of graphene~\cite{balakrishnan2013colossal} improves the SOC in graphene at the cost of  introducing disorder and reducing the graphene's mobility. An alternate technique is interfacing graphene with two-dimensional transition metal dichalcogenides (TMDC) having a high SOC~\cite{PhysRevLett.119.146401, gmitra2009band, PhysRevB.96.041409, avsar2014spin, PhysRevX.6.041020, wang2015strong, safeer2019room, PhysRevLett.121.127703, garcia2017spin,han2014graphene,wakamura2018strong,ghiasi2017large,offidani2018microscopic,luo2017opto,ghiasi2019charge,benitez2020tunable}.
	
	Before one conceives increasingly complex graphene/TMDC heterostructures and considers their potential applications, it is  imperative to understand the impact of the proximity of TMDC on the electronic properties of graphene.  Prominent amongst these are the breaking of inversion symmetry, breaking of sub-lattice symmetry, and hybridization of the \textit{d}-orbitals of the heavy element in TMDC with the \textit{p}-orbitals of SLG, leading to strong SOC in SLG. This proximity-induced SOC in SLG has three primary components, all of which contribute to spin splitting of the bands -- (i) valley-Zeeman (also called Ising) term, which couples the spin and valley degrees of freedom, (ii) Kane-Mele term~\cite{PhysRevLett.95.226801,PhysRevLett.95.146802} which couples the spin, valley and sublattice components and opens a topological gap at the Dirac point~\cite{C7CS00864C,PhysRevB.99.205407}, and (iii) Rashba term~\cite{gmitra2016trivial} which couples the spin and sublattice components. 
	
	In the presence of a strong Ising SOC, the electronic band dispersion of graphene is predicted to be spin-split~\cite{gmitra2016trivial,cysne2018quantum,PhysRevLett.119.146401,doi:10.1021/acs.nanolett.7b03604,PhysRevLett.119.206601,PhysRevB.92.155403}, as was observed recently in bilayer graphene/\ch{WSe2} heterostructures~\cite{island2019spin,tiwari2021electric}. Consequences of this induced SOC include the appearance of helical edge modes and quantized conductance in the absence of a magnetic field in bilayer graphene~\cite{tiwari2020observation} and of weak antilocalization~\cite{wang2015strong},   and Hanle precession in  SLG~\cite{PhysRevB.97.045414,PhysRevLett.119.196801,PhysRevLett.121.127702,PhysRevX.6.041020,Benitez2018,PhysRevB.97.075434,PhysRevB.97.045414,PhysRevLett.121.127703}. Despite these advances, a quantitative study of the effect of a strong SOC on the electronic energy band dispersion of SLG is lacking.
	
	In this research article, we report the results of our studies of quantum oscillations in high-mobility heterostructures of SLG and trilayer \ch{WSe2}. Careful analysis of the oscillation frequencies shows spin-splitting of the order of $\sim 5$~meV for both the valence band (VB) and the conduction band (CB). We find that the bands remain linear down to at least 70~meV (corresponding to $n\sim 2 \times 10^{11} \mathrm{cm^{-2}}$). Close to zero energy, the lower energy branches of the CB and the VB overlap, leading to band inversion and opening of a band gap in the energy dispersion of SLG. We fit our data using a theoretical model that establishes that, to the zeroth-order, the magnitude of the spin-splitting of the bands and that of the band gap are determined by only the valley-Zeeman and Rashba spin-orbit interactions.

	\section{Results}
	
	\subsection{Experimental Observations}
	
	\begin{figure}[t!]
		\begin{center}
			\includegraphics[width=\textwidth]{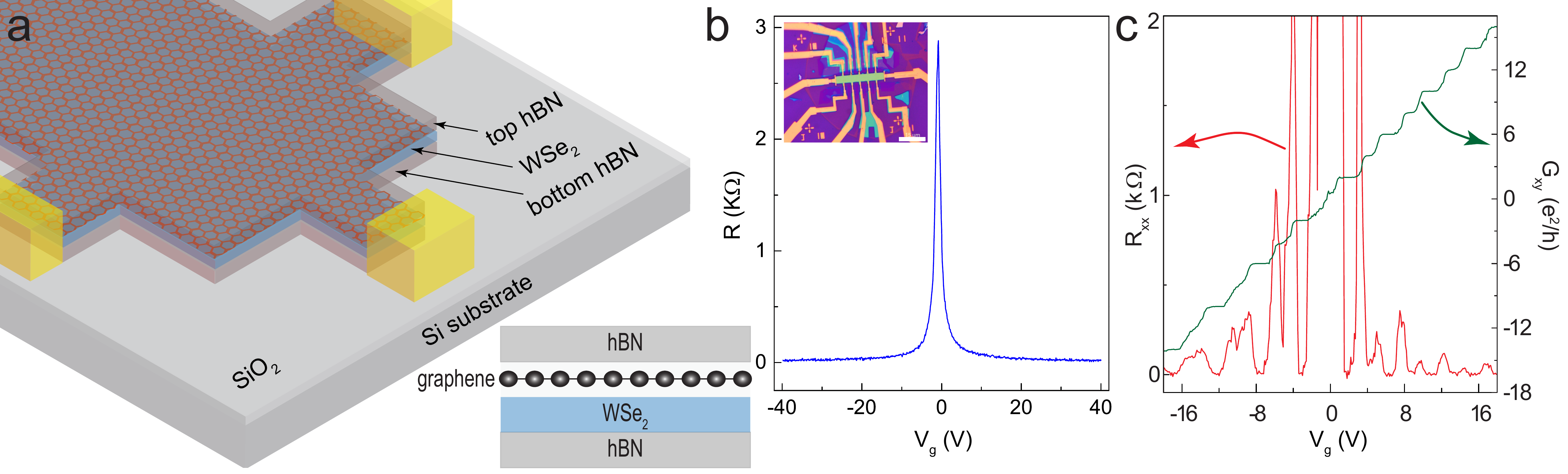}
			\caption{\textbf{Transport measurements in hBN encapsulated SLG/WSe$_2$ heterostructures.} (a) A schematic of the device. The bottom right corner shows the sequence of all the layers of the heterostructure. (b) Four-probe resistance of the device plotted as a function of the back-gate voltage $V_\mathrm{\rm bg}$; the data were collected at 20~mK. An optical image of the device is shown in the inset. The scale bar in the figure is 15~$\mathrm{\mu}$m. (c) Plots of longitudinal resistance $R_\mathrm{xx}$ (left-axis) and the Hall conductance $G_\mathrm{xy}$ (right-axis) versus $V_\mathrm{\rm bg}$ in the quantum Hall regime. The measurements were done at 20~mK in the presence of a 3~T perpendicular magnetic field.			\label{fig:figure1} }
		\end{center}
	\end{figure}		
	Heterostructures of single-layer graphene and trilayer \ch{WSe2}, encapsulated by hexagonal boron nitride (hBN) (see device schematic Fig.~\ref{fig:figure1}(a)) of thickness~$\sim$~20-30~nm, were fabricated using  dry transfer technique~\cite{pizzocchero2016hot,wang2013one}. One-dimensional Cr/Au electrical contacts were created by standard nanofabrication techniques -- note that this method completely evades contacting the \ch{WSe2} thus avoiding parallel channel transport (see Supplementary Information for details). Electrical transport  measurements were performed using a low-frequency ac lock-in technique in a dilution refrigerator at the base temperature of 20~mK unless specified otherwise. Multiple devices of SLG/\ch{WSe2} were studied, and the data from all of them were qualitatively very similar. All the data presented here are from a device labeled B9S6. The data for two other similar devices are presented in the Supplementary Information.  The extracted impurity density from the four-probe resistance of the device as a function of gate voltage (see Fig.~\ref{fig:figure1}(b)) was $\sim 2.2\times 10^{10}$~cm$^{-2}$, and the mobility was  $\sim$140,000~cm$^2$V$^{-1}$s$^{-1}$. The four-probe resistance response as a function of the gate voltage were identical for different measurement configurations (see Supplementary Information), indicating that the fabricated device is spatially homogeneous. Fig.~\ref{fig:figure1}(c) shows the quantum Hall data at 3~T -- the presence of plateaus at $\nu=\pm \ 2,~\pm \ 6,~\pm \ 10$ confirms it as SLG. Along with the signature plateaus of SLG, one can see a few of the broken symmetry states appearing already at 3~T, confirming it to be a high-quality device.
	
		\begin{figure}[t!]
		%\begin{center}
		\includegraphics[width=0.65\textwidth]{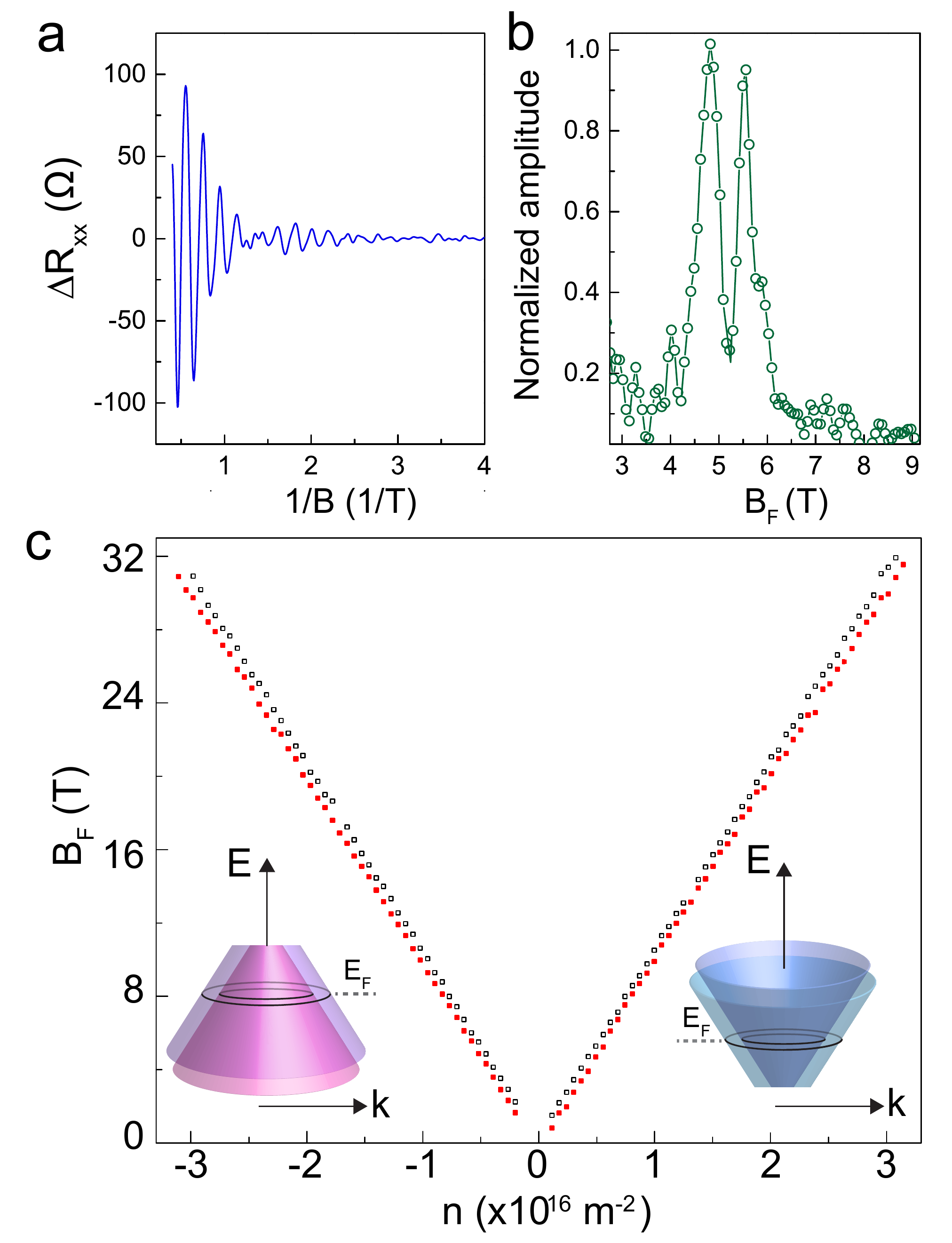}
		\caption{\textbf{Shubnikov-de Haas(SdH) oscillations} (a) Plot of the Shubnikov-de Haas oscillations measured at $V_\mathrm{bg} = -9$~V and $T = 20$~mK. (b) The corresponding FFT spectra showing two distinct peaks. (c) Charge carrier density ($n$) dependence of the frequency of SdH oscillations. The inset shows the schematics of the spin split conduction and valance bands.
			\label{fig:figure2}}
		%\end{center}
	\end{figure}		
	
	Representative data of the Shubnikov-de Haas (SdH) oscillations measured at 20~mK are plotted in Fig.~\ref{fig:figure2}(a). In addition to the expected decay of the amplitude of the oscillations with increasing $1/\mathrm{B}$, we observe the presence of beating, implying two closely spaced frequencies. The fast Fourier transforms (FFT) of the data  Fig.~\ref{fig:figure2}(b) show that this indeed is the case. We find similar splitting in the SdH oscillation frequency in all the SLG/\ch{WSe2} devices studied by us -- the data for two additional similar devices are presented in the  Supplementary Information. There may be a legitimate concern that the observed beating can be caused by device inhomogeneities which lead to different charge carrier density in different regions of the graphene channel. We rule out this artifact from measurements of the four-probe resistance and SdH oscillations in multiple contact configurations -- we find that the data are identical in each case (see Supplementary Information). 
	
	Recall that the SdH oscillation frequency, $B_\mathrm{F}$ is directly related to the cross-sectional area $A\left(\mathrm{\bold{k}}\right)$ at the Fermi energy by the relation $B_\mathrm{F}=\hbar A\left(\mathrm{\bold{k}}\right)/2\mathrm{\pi e}$~\cite{novoselov2005two}. For an isotropic dispersion in which the Fermi energy $E_\mathrm{F}$ is a function of $k_\mathrm{F} = \sqrt{k_\mathrm{x}^2 + k_\mathrm{y}^2}$ (where $(k_\mathrm{x},k_\mathrm{y})$ are defined with respect to one of the Dirac points, $K$ or $K'$, of the SLG), the cross-sectional area of the Fermi surface is given by $A(\mathrm{\bold{k}}) = \mathrm{\pi} k_\mathrm{F}^2$. Fig.~\ref{fig:figure2}(c) shows the charge carrier density ($n$) dependence of  $B_\mathrm{F}$. The appearance of two closely spaced frequencies at all $n$ (or $E_\mathrm{F}$) implies that for each value of the Fermi energy, there are two distinct values of $k_\mathrm{F}$. This is a direct proof of the energy splitting of both the CB and the VB of the SLG.

	\begin{figure}[t!]
		%\begin{center}
		\includegraphics[width=0.75\textwidth]{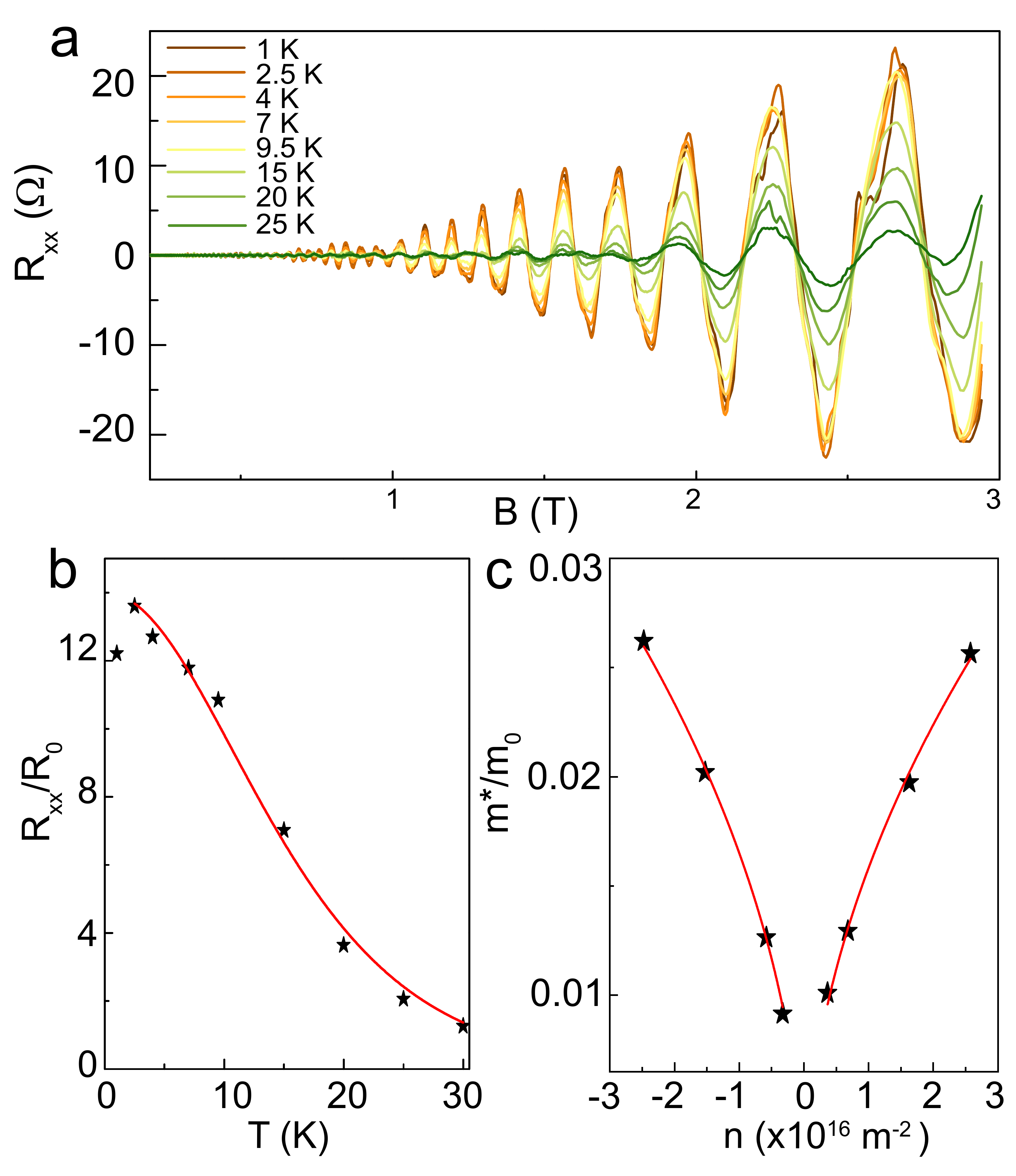}
		\caption{\textbf{Temperature dependent SdH oscillations and effective mass extraction} (a) SdH oscillations for different temperatures measured at  $V_\mathrm{bg}=25$~V. (b) The normalized amplitude of the SdH oscillations as a function of temperature (black stars). The solid red line is the fit to Eq.~\ref{rhoxx} used to extract the effective mass. (c) Plot of the effective mass $m^{\ast}$ versus the charge carrier density $n$ (black stars).  The solid red curve is the fit used to extract the relation between $m^{\ast}$ and $n$.	
			\label{fig:figure3}}
		%\end{center}
	\end{figure}

	From the temperature dependence of the amplitude of the SdH oscillations (Fig.~\ref{fig:figure3}), we extracted the effective charge carrier mass $m^{\ast}$, using the the Lifshitz-Kosevich relation~\cite{kuppersbusch2017modifications,lifshitz1956theory}:
	\begin{equation}
		\frac{\Delta R_\mathrm{xx}}{R_\mathrm{0}}\propto\frac{2\mathrm{\pi}^2k_\mathrm{B}Tm^{\ast}/\hbar \mathrm{eB}}{sinh\left(2\mathrm{\pi}^2k_\mathrm{B}Tm^{\ast}/\hbar \mathrm{eB}\right)}, \label{rhoxx}
	\end{equation}
	Here, $R_\mathrm{0}$ the longitudinal resistivity at $\mathrm{B}=0$. On fitting the effective mass $m^{\ast}$ versus $n$ using the relation $m^\ast=\hbar(\sqrt{\mathrm{\pi}}/v_\mathrm{F})n^\alpha$ ~\cite{neto2009electronic} (see Supplementary Information), we obtain $\alpha=0.5\pm 0.02$ and Fermi velocity $v_\mathrm{F} = 1.29 \pm 0.04 \times {10}^{6}$~ms$^{-1}$. The value of $\alpha$ being $0.5$ establishes the dispersion relation between energy and momentum in SLG on \ch{WSe2} to be linear~\cite{novoselov2005two}.

	Fig.~\ref{fig:figure4}(a) are the resultant plots of $E$ versus $k$ for both CB and the VB from our experimental data. Note that our experimental data extends down to $\approx 30$~meV (corresponding to ${n} \approx3.9 \times10^{10}$~$\mathrm{cm^{-2}}$). Below this number density, the SdH oscillations are not resolvable -- presumably due to the dominance of charge puddles on the electrical transport of SLG in this energy range.

	We observe that, on extrapolating the $E-k$ plots to $E=0$, the low-energy branches of the spin-split bands of both the CB and the VB bands enclose a finite area in the $k$-space at $E=0$. This leads us to expect that there will be an overlap between the lower branches of the CB and VB, ultimately leading to band inversion near the $K$ (and $K'$) points. A verification of this assertion requires further measurements in extremely high quality devices that will allow measurements of SdH oscillations near $E=0$.
	
	To summarize our experimental observations, we have quantified the spin-splitting of the energy bands in SLG in proximity to \ch{WSe2} and mapped out the dispersion relation of the spin-split bands of  SLG.  We find that till a certain energy, the dispersion remains linear; below this energy scale, we observe a deviation from linearity.

	\subsection{Theoretical calculations}

	\begin{figure}[t!]
		%\begin{center}
		\includegraphics[width=\textwidth]{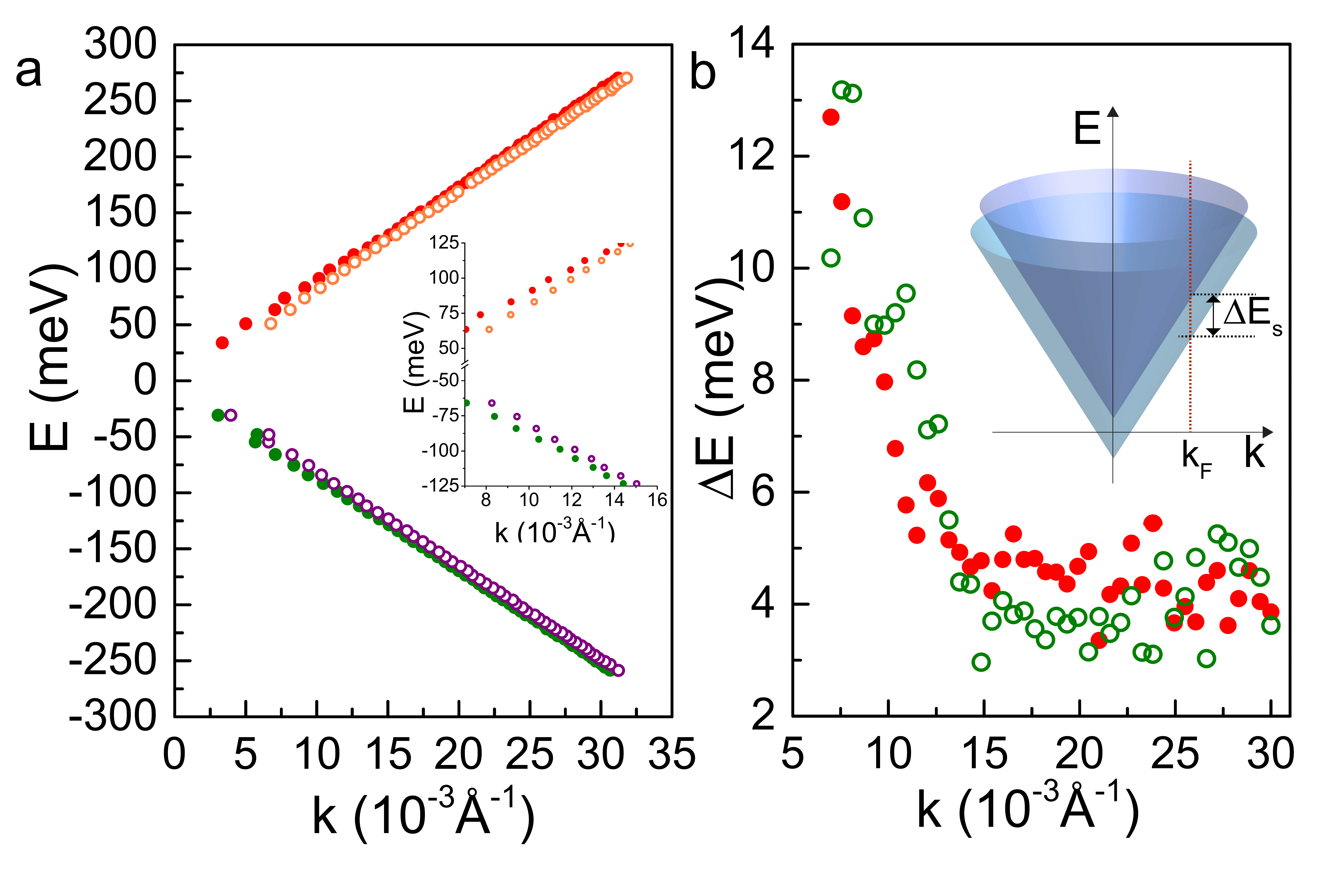}
		\caption{\textbf{Experimentally obtained dispersion relation} (a) Dispersion relation for SLG on \ch{WSe2} as extracted from our SdH measurements. The inset shows the zoomed-in plot near the Dirac point. (b) Plot of the magnitude of the energy difference between the two spin-split bands, $\Delta E_\mathrm{s}$ versus $k$ for the valence band (green open circles) and the  conduction band (red filled circles). The inset shows a schematic of the band-splitting in the CB -- $\Delta E_\mathrm{s}$ is the spin-splitting at $k_\mathrm{F}$.
			\label{fig:figure4}}
		%\end{center}
	\end{figure}

	Using the experimental data, we fit a theoretical model to obtain the dispersion relation close to the Dirac points. The continuum Hamiltonian near the Dirac points for SLG with WSe$_2$ has the following terms (see, for instance, Ref.~\cite{gmitra2016trivial}):
	\begin{eqnarray}
		H& =& \hbar v_\mathrm{F} (\eta k_\mathrm{x} \sigma_\mathrm{x} + k_\mathrm{y} \sigma_\mathrm{y})+ \Delta \sigma_\mathrm{z} + \lambda_\mathrm{KM} \eta S_\mathrm{z} \sigma_\mathrm{z}  + \lambda_\mathrm{VZ} \eta S_\mathrm{z} +  \lambda_\mathrm{R} (\eta S_\mathrm{y} \sigma_\mathrm{x} - S_\mathrm{x}\sigma_\mathrm{y}) \nonumber \\ &&+ \frac{\sqrt{3}a}{2} [\lambda_\mathrm{PIA}^{A} (\sigma_\mathrm{z} +\sigma_\mathrm{0}) + \lambda_\mathrm{PIA}^{B}(\sigma_\mathrm{z} -\sigma_\mathrm{0})] (k_\mathrm{x} S_\mathrm{y} - k_\mathrm{y} S_\mathrm{x}) \label{equation2}
	\end{eqnarray}
	In Eq.~\ref{equation2}, the Pauli matrices $\sigma_\mathrm{i}$ and the $S_\mathrm{i}$ represent the sublattice and spin degrees of freedom, respectively. The first term denotes the linear dispersion near the Dirac points, where $v_\mathrm{F}$ is the Fermi velocity, $k_\mathrm{x}$ and $k_\mathrm{y}$ are the momenta with respect to the Dirac point, and $\eta = \pm 1$ denotes the valleys $K$ ($K'$) respectively
	(We note that $\hbar v_\mathrm{F}=3ta/2$, where $t$ is the nearest-neighbor hopping amplitude and the nearest neighbour carbon carbon distance $a$ is 1.42 \AA). The second term represents a sublattice potential of strength $\Delta$. We have considered the four possible spin-orbit couplings:
	(i) Kane-Mele SOC with strength $\lambda_\mathrm{KM}$, (ii) valley-Zeeman SOC with strength $\lambda_\mathrm{VZ}$, (iii) Rashba SOC with strength $\lambda_R$,  and (iv) pseudo-spin asymmetric SOC with strengths $\lambda_\mathrm{PIA}^{A}$ and $\lambda_\mathrm{PIA}^{B}$ for sublattices A and B respectively. 
	
	\begin{figure}[t!]
	%\begin{center}
	\includegraphics[width=\textwidth]{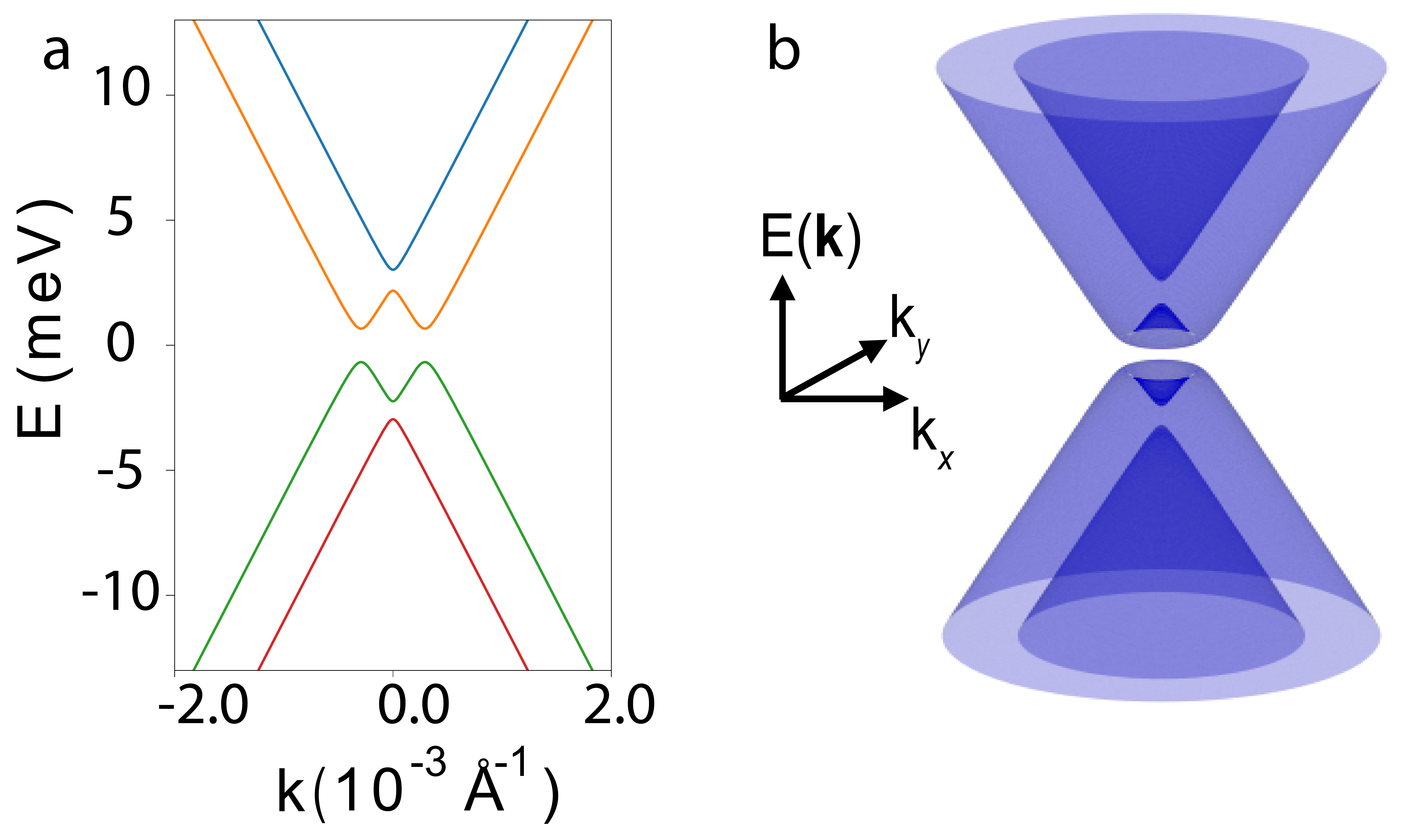}
	\caption{\textbf{Theoretically calculated dispersion relation} (a) A 2D plot of the energy dispersion relation calculated with  $v_\mathrm{F}= 1.286 \times 10^6$ $\mathrm{m s^{-1}}$, and $\theta = 13 ^{\circ}$ ($\lambda_\mathrm{VZ} \simeq \pm 2.45$~meV and $\lambda_\mathrm{R}= \pm 0.56$~meV). The value of $\lambda_\mathrm{R}$ is taken from Ref.~\cite{gmitra2016trivial}. The value of $\lambda_\mathrm{VZ}$ comes from fitting the experimental spin split band gap energy.  Further, the other parameters in this graph have  been set as follows: $\Delta= 0.54$~meV, $\lambda_\mathrm{KM}= 0.03$~meV, $\lambda_\mathrm{PIA}^{A}= -2.69$~meV and $\lambda_\mathrm{PIA}^{B}= -2.54$~meV. (b) The 3D dispersion of this model for the same set of parameters as in (a).
		\label{fig:figure5}}
	%\end{center}
\end{figure}	

	Since this Hamiltonian results in the same dispersion at both the valleys, we only consider the case $\eta = +1$ ($K$ point). The Hamiltonian in Eq.~\ref{equation2} is invariant under a simultaneous rotation of $(k_\mathrm{x},k_\mathrm{y})$, $(\sigma_\mathrm{x},\sigma_\mathrm{y})$ and $(S_\mathrm{x},S_\mathrm{y})$ by the same angle; this implies that the dispersion is isotropic in momentum space, and it is sufficient to take $k_\mathrm{x} = k$ and $k_\mathrm{y} =0$. The data for the four bands shown in Fig.~\ref{fig:figure4}(a) are fitted to the Hamiltonian with  $t, ~\Delta, ~\lambda_\mathrm{KM}, ~\lambda_\mathrm{VZ}$ and $\lambda_\mathrm{R}$ as the fit parameters. The best fit gives $t= 3979.10\pm 3.99$ meV implying a large Fermi velocity in this device of $1.286 \times 10^6$~ms$^{-1}$ (compared to about $0.86 \times 10^6$~ms$^{-1}$ in pristine SLG~\cite{hwang2012fermi}). The parameters in the Hamiltonian which give the spin-split band gap in both conduction and valence bands are  $\lambda_{VZ}$ and $\lambda_R$. We find that the best fit gives the values of $\lambda_\mathrm{VZ}$ and $\lambda_\mathrm{R}$ to lie on a circle of radius $2.51$~meV, such that 
	\begin{equation} \label{equation3}
		\lambda_\mathrm{VZ}= 2.51 ~\cos\theta ~~\rm{meV}, ~~~{\rm and}~~~ 
		\lambda_\mathrm{R} = 2.51 ~\sin\theta ~~\rm{meV}.
	\end{equation}
	
	\noindent where $\theta$ can take any value from $0$ to $2\mathrm{\pi}$, and $\Delta = \lambda_\mathrm{KM} =0$. Eq.~\ref{equation3} can be understood by looking at the first-order perturbative effect of the valley-Zeeman and Rashba terms in the Hamiltonian. Taking $H_\mathrm{0} = \hbar v_\mathrm{F} k \sigma_\mathrm{x}$ and the perturbation $V=\lambda_\mathrm{VZ} S_\mathrm{z} + \lambda_\mathrm{R} (S_\mathrm{y} \sigma_\mathrm{x} - S_\mathrm{x} \sigma_\mathrm{y}),$ we find that the zeroth order spin-degenerate dispersion $E_\mathrm{0} = \pm \hbar v_\mathrm{F} k$ in the positive and negative energy bands receives first-order corrections given by
	\begin{equation} \label{equation4}
		E_\mathrm{1,\pm} ~=~ \pm \sqrt{\lambda_\mathrm{VZ}^2 +\lambda_\mathrm{R}^2} ~=~ \pm 2.51 ~{\rm meV}
	\end{equation}
	for both the bands, thus giving the general relation in Eq.~\ref{equation3}. This gives a gap equal to twice $2.51$~meV which fits the experimentally observed value of $\sim 5$~meV. The two extreme cases are given by $\theta= 0$ with only valley-Zeeman SOC and $\theta= \mathrm{\pi/2}$ with only Rashba SOC. 	The overall magnitude of effective SOC of $2.51$~meV agrees well with previous reports~\cite{PhysRevB.97.075434, wang2015strong, C7CS00864C, gmitra2016trivial}. 
	
	The band dispersion for higher energies ($E >5$~meV) remains unaffected for any combination of $\lambda_\mathrm{VZ}$ and $\lambda_\mathrm{R}$ which satisfies $\sqrt{\lambda_\mathrm{VZ}^2 + \lambda_\mathrm{R}^2} = 2.51$~meV. However, the relative magnitudes of $\lambda_\mathrm{VZ}$ and $\lambda_\mathrm{R}$ modifies the lower energy band dispersion ($E<5$~meV) -- a region inaccessible in our experiments. To evaluate the value of $\lambda_\mathrm{VZ}$ and $\lambda_\mathrm{R}$ explicitly, one needs the value of the band gap at $E=0$. As is well known, the presence of finite impurity density ($n_\mathrm{0} \approx {2.2} \times10^{10}$~$\mathrm{cm^{-2}}$ for this device) leads to the dominance of charge puddles on the electrical transport of SLG at the Dirac point making it extremely difficult to make an accurate estimate of such a small energy gap. In the absence of such information, we take the theoretically predicted value of $\lambda_\mathrm{R} = 0.56$~meV~\cite{gmitra2016trivial} which corresponds to $\theta= 13 ^{\circ} $ in Eq. ~\ref{equation3}. This yields the strength of the valley-Zeeman term to be $\lambda_\mathrm{VZ} = 2.45$~meV and a maximum expected band gap of $3.3$~meV for $\lambda_\mathrm{so} \sim 2.51$~meV. The resulting dispersion is plotted in Fig.~\ref{fig:figure5}(a). In generating this plot, we have used $\Delta= 0.54$~meV, $\lambda_\mathrm{km}= 0.03$~meV, $\lambda_\mathrm{PIA}^{A}= -2.69$~meV and $\lambda_\mathrm{PIA}^{B}= -2.54$~meV~\cite{gmitra2016trivial}. Fig.~\ref{fig:figure5}(b) further shows the 3-dimensional plot of the energy dispersion for this model. 
	
	\section{Discussions}
	
	Coming to the role of the magnetic field in the extracted energy dispersion relation, the SdH oscillations were studied at a magnetic field of the order of 1~T. This field gives a very small Zeeman energy of the order of $0.08~$meV  in the energy. Since the experimental data points are quite far from $E=0$, we can ignore the magnetic field effect in the Hamiltonian. Further, the fitting does not give us the values of $\Delta$, $\lambda_\mathrm{KM}$ , $\lambda_\mathrm{PIA}^{A}$, and $\lambda_\mathrm{PIA}^{B}$  with any certainty. While the $PIA$ terms do not alter the dispersion in our region of interest, the other two parameters, $\Delta$ and $\lambda_\mathrm{KM}$  will open up a gap at the two values of the energy lying at $k=0$. The effect of the Kane-Mele term in this model has been further discussed in the Supplementary Information.  However, we note that the presence of $\Delta$, $\lambda_\mathrm{KM}$, $\lambda_\mathrm{PIA}^{A}$ and $\lambda_\mathrm{PIA}^{B}$ does not alter the spin-split band gap of 5~meV observed between the bands away from zero energy.

	All  previous theoretical and experimental studies note that the valley-Zeeman $\lambda_\mathrm{VZ}$ and the Rashba $\lambda_\mathrm{R}$ are the major spin-orbit coupling (SOC) terms for graphene/TMDs~\cite{PhysRevB.97.075434, wang2015strong, C7CS00864C, gmitra2016trivial}. These two terms by themselves give a constant energy gap between the spin-split bands. The other relevant spin-orbit coupling terms are $\lambda_\mathrm{PIA}$ and $\lambda_\mathrm{KM}$. The $\lambda_\mathrm{PIA}$ terms are negligibly small and do not alter the band dispersion in the region of interest.  On the other hand, we find that including a very large  $\lambda_\mathrm{KM}$ and $\Delta$ terms $\sim 20$~meV in the standard theoretical models can give a dispersion where the energy gap between the two spin-split bands increases as one approaches the Dirac point. However, such large $\lambda_\mathrm{KM}$ and $\Delta$ might not be reasonable and have not been reported to date. To the best of our knowledge, there is no consistent theoretical understanding of the increase in energy gap between the spin-split bands on approaching the Dirac point; we leave this as an open question to be explored in near future.

	Finally, a comment on the relative magnitudes of $\lambda_\mathrm{VZ}$ and $\lambda_\mathrm{R}$: The spin relaxation mechanism in graphene/TMDC heterostructures is extraordinary. It relies on intervalley scattering and can only occur in materials with spin-valley coupling. In such systems, the lifetime $\tau$ and relaxation length $\lambda$ of spins pointing parallel to the graphene plane ($\tau_\mathrm{s}^{\parallel}$, $\lambda_\mathrm{s}^{\parallel}$) can be markedly different from those of spins pointing out of the graphene plane ($\tau_\mathrm{s}^{\perp}$, $\lambda_\mathrm{s}^{\perp}$).  Realistic modeling of experimental studies   indicate that the spin lifetime anisotropy ratio  $\zeta=\tau_\mathrm{s}^{\parallel}/\tau_\mathrm{s}^{\perp}={{{(\lambda}_\mathrm{s}^{\perp}/\lambda}_\mathrm{s}^{\parallel})}^2$ can be as large as a few hundred in the presence of intervalley scattering~\cite{PhysRevB.97.045414,Benitez2018, PhysRevLett.119.206601}. Recall that the $\lambda_\mathrm{VZ}$ provides an out-of-plane spin-orbit field and affects the in-plane spin relaxation time, $\tau_\mathrm{s}^{\parallel}$. On the other hand, $\lambda_\mathrm{R}$ generates an in-plane spin-orbit field and is relevant for determining $\tau_\mathrm{s}^{\perp}$~\cite{Benitez2018}. The large spin lifetime anisotropy ratio ($\tau_\mathrm{s}^{\parallel}/\tau_\mathrm{s}^{\bot} \gg 1$) seen both from experiments and theory~\cite{PhysRevB.97.045414,Benitez2018, PhysRevLett.119.206601} show that the value of $\lambda_\mathrm{VZ}$ can indeed be significantly larger as compared to $\lambda_\mathrm{R}$.

	In conclusion, we have experimentally determined the band structure of single-layer graphene in the presence of proximity-induced SOC. We find both the VB and the CB spin-split with a spin-energy gap of $\sim 5$~meV; the splitting increases as one approaches the Dirac point. There are strong indications of overlap of the lower energy branches of the conduction and the valence bands. We also provide  precise values of the spin splitting energy, the Fermi velocity and the effective mass of charge carriers in graphene/WSe$_2$ hetrostructures.  Theoretical modeling of the data establishes that the band dispersion near the Dirac point and the magnitude of the spin-splitting are determined primarily by large valley-Zeeman (Ising) SOC and small Rashba SOC. Our work raises the strong possibility that in this system, the transport properties near the Dirac point are dominated by charge carriers of a single spin component, making this system a potential platform for realizing spin-dependent transport phenomena, such as quantum spin-Hall and spin-Zeeman Hall effects.

	\section*{Methods}
	\subsection*{Device Fabrication}
	
	The SLG, \ch{WSe2}, and hBN flakes were obtained by  mechanical exfoliation on \ch{SiO2}/\ch{Si} wafer using scotch tape from the corresponding bulk crystals. The thickness of the flakes was verified from Raman spectroscopy. Heterostructures of SLG and \ch{WSe2}, encapsulated by single-crystalline hBN flakes of thickness $\sim$20-30~nm was fabricated by dry transfer technique using a home-built transfer set-up consisting of high-precision XYZ-manipulators. The heterostructure was then annealed at $250^\circ$C for 3 hours. Electron beam lithography followed by reactive ion etching (where the mixture of \ch{CHF3} and \ch{O2} gas were used with flow rates of 40 sccm and 4 sscm, respectively, at a temperature of 25$^\circ$C at the RF power of 60~W) was used to define the edge contacts. The electrical contacts were fabricated by depositing Cr/Au (5/60~nm) followed by lift-off in hot acetone and IPA. 
	\subsection*{Measurements}
	All electrical transport  measurements were performed using a low-frequency AC lock-in technique in a dilution refrigerator (capable of attaining a lowest temperature of 20~mK and maximum magnetic field of 16~T).

	\textbf{Data availability}

	The authors declare that the data supports the findings of this study are available within the main text and its supplementary Information. Other relevant data are
	available from the corresponding author upon reasonable request.

	\textbf{Acknowledgment}: 
	
	The authors acknowledge fruitful discussions with Saurabh Kumar Srivastav and Ramya Nagarajan and facilities in CeNSE, IISc. AB acknowledges funding from DST FIST program and DST (No. DST/SJF/PSA01/2016-17). DS acknowledges funding from SERB (JBR/2020/000043). K.W. and T.T. acknowledge support from the Elemental Strategy Initiative conducted by the MEXT, Japan (Grant Number JPMXP0112101001) and  JSPS KAKENHI (Grant Numbers JP19H05790 and JP20H00354). DSN  thanks DST for Woman Scientist  fellowship (WOS-A) (Grant No.SR/WOS-A//PM-98/2018).
	
	\textbf{Competing interests:}
	
	The authors declare no Competing Financial or Non-Financial Interests.
	
	\textbf{Author Contributions}
	
	P.T. and A.B. conceived the idea of this research. P.T. and D.S.N. fabricated the devices; P.T., M.K.J. and A.B. performed the measurements; A.U. and D.S. provided theoretical support; K.W and T.T. grew the material; P.T., M.K.J. and A.B. did the experimental data analysis. P.T., A.U. and A.B. co-wrote the manuscript. All authors discussed the results and
	commented on the manuscript.

\appendix
\renewcommand\thefigure{S\arabic{figure}} 

\newpage

\section*{Supplementary Materials}

\section{Device Fabrication and characterization}

We fabricated heterostructures of single-layer graphene (SLG) and trilayer~\ch{WSe2}, encapsulated by single-crystalline hBN flakes of thickness $\sim$20-30~nm. The SLG, \ch{WSe2}, and hBN flakes were obtained by  mechanical exfoliation on \ch{SiO2}/\ch{Si} wafer using scotch tape from the corresponding bulk crystals. The thickness of the flakes was verified both from optical contrast under an optical microscope and Raman spectroscopy.

\begin{figure*}[t]
	\begin{center}
		\includegraphics[width=\textwidth]{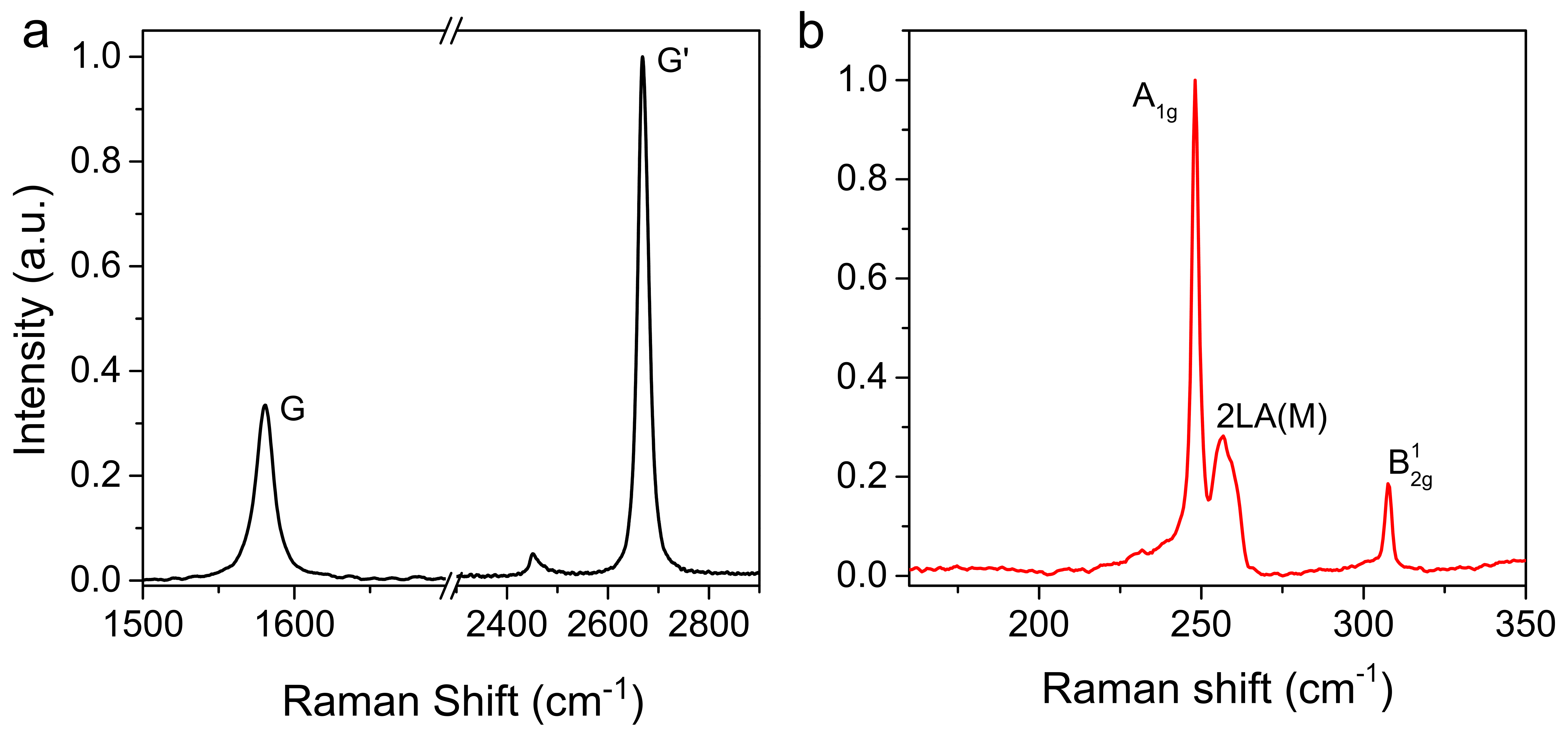}
		\caption{\textbf{Raman spectra} Room temperature Raman spectra for (a) the single-layer graphene and (b)  the \ch{WSe2} flake.}
		\label{fig:Raman}
	\end{center}
\end{figure*}

The Raman data for SLG and trilayer \ch{WSe2} flake are shown in Figure.~\ref{fig:Raman}(a) and (b) respectively. For the graphene, the high intensity of the Lorentzian G' peak confirms it to be a single-layer. 
The flakes were formed into a heterostructure using dry transfer technique~\cite{pizzocchero2016hot} using a home-built transfer set-up consisting of high-precision XYZ-manipulators, the entire process being performed under an optical microscope.Briefly, the hBN was first picked up using a Polycarbonate (PC) film at 90$^o$C. This combination was then used to pick up the SLG  followed by \ch{WSe2} and hBN.
The prepared stack was transferred on a clean Si/\ch{SiO2} wafer at 180$^o$C and cleaned using chloroform to remove the PC, and this was followed by cleaning with acetone and isopropyl alcohol.The heterostructure was then annealed at $250^\circ$C for 3 hours.

\begin{figure*}[t!]
	\begin{center}
		\includegraphics[width=0.8\textwidth]{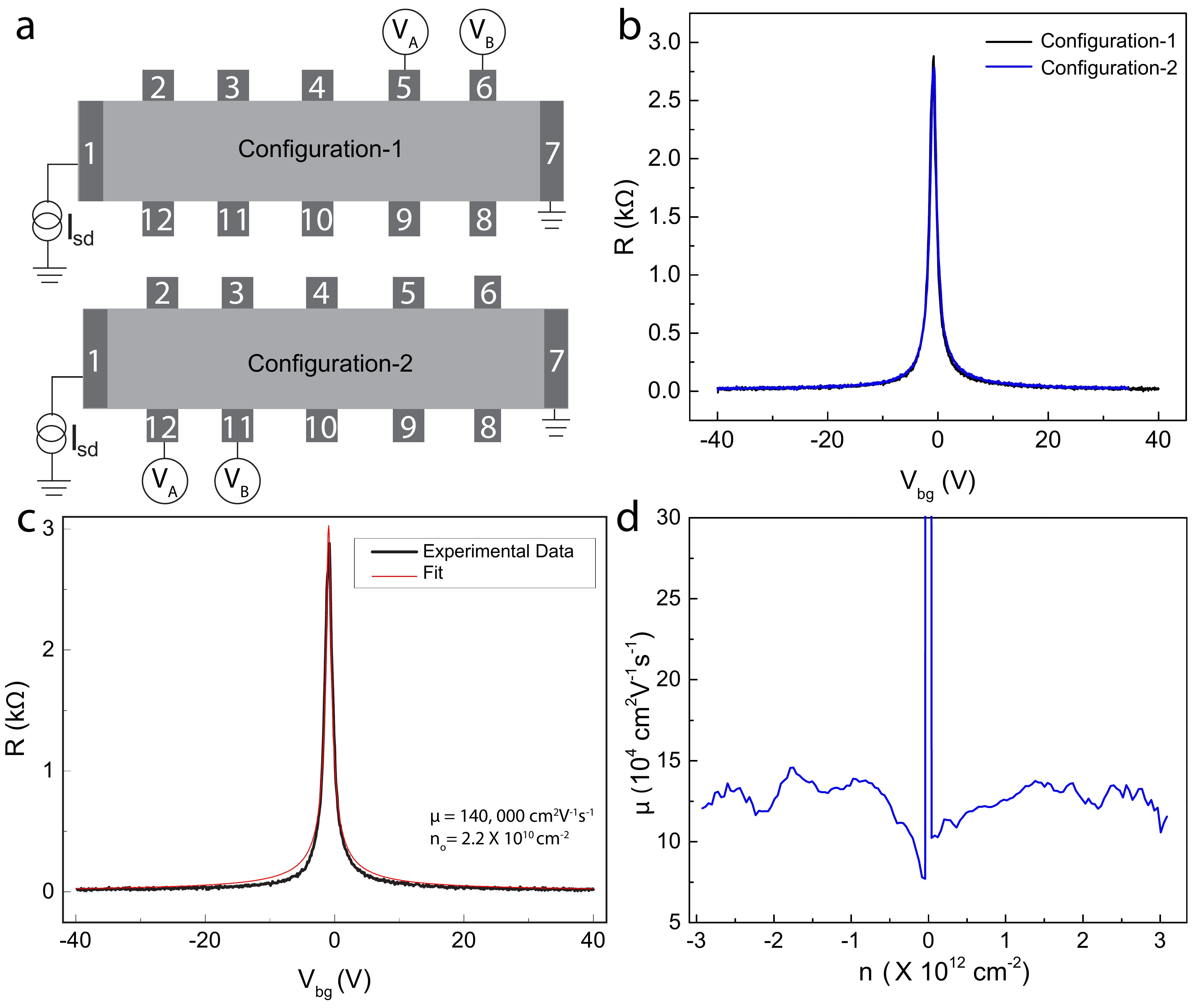}
		\caption{\textbf{Resistance measurements in different configurations.} (a) Measurement configurations for the  data shown in (b). (b) Four probe resistance plotted as a function of gate voltage at 20~mK, the blue and the black lines correspond to two different configuration shown in (a). (c)  Four probe resistance plotted as a function of gate voltage at 20~mK (shown by a black curve) - the red curve is the fit to the data for the extraction of mobility ($\mu$) and impurity density ($n_\mathrm{0}$). The $\mu$ of the device was $\sim 140,000$~cm$^2$V$^{-1}$s and $n_\mathrm{0}$ was $2.2 \times{10}^{10}$~cm$^{-2}$. (d) Plot of  mobility ($\mu$) vs charge carrier density ($n$) at 20~mK.}
		\label{fig:mobility}
	\end{center}
\end{figure*}

Electron beam lithography was used to define the edge contacts. The edge contacts were made by reactive ion etching (where the mixture of \ch{CHF3} and \ch{O2} gas was used with flow rates of 40 sccm and 4 sscm, respectively, at a temperature of 25$^\circ$C at the RF power of 60~W)~\cite{Wang614}. The electrical contacts were finally created by depositing Cr/Au (5/60~nm) followed by lift-off in hot acetone and IPA. Cr/Au was chosen as it forms a very high quality ohmic contact with graphene~\cite{Wang614}, but at the same time it does not form any contact with \ch{WSe2} due to high Schottky barrier and large difference in work functions~\cite{smyth2017wse2, wang2016does}. Finally, the device was etched into a Hall bar geometry. An optical image of the final device is shown in the main text (inset of Fig.~\ref{fig:figure1}(b)).

To estimate the impurity density ($n_\mathrm{0}$) and field-effect mobility ($\mu$) of the device, the gate-voltage dependent resistance data were fitted by the equation
\begin{equation}
	R={2R}_\mathrm{c}+L/W\mathrm{e}\mu\sqrt{n_\mathrm{0}^2+\left(V_\mathrm{bg}-{\ V}_\mathrm{d}\right)^2C^2/\mathrm{e}^2},
	\label{eqn_Rvg}
\end{equation}
where $R_\mathrm{c}$ is the contact resistance, $L$ and $W$ are the channel length and width, respectively, $C$ is the capacitance per unit area, and $V_\mathrm{d}$ is the value of the back-gate voltage  at the Dirac point. Supplementary Figure~\ref{fig:mobility} (b)is the plot of four probe resistance versus gate voltage {in two different measurement configurations (see Supplementary Figure~\ref{fig:mobility}(a)). One can see that the data for the two contact configurations are identical indicating that there is no local shift in the Dirac point in the graphene channel. This shows that the fabricated device is homogeneous and  confirms that the observed beating in the SdH oscillation does not arise from spatial number density inhomogeneity.}

{Supplementary Figure~\ref{fig:mobility}(c) shows the mobility extraction fitting curve (red)} using Eq.~\eqref{eqn_Rvg}. The extracted $n_\mathrm{o}$ was $\sim2.2 \times 10^{10}$ cm$^{-2}$, and $\mu$ was found to be nearly $\sim$140,000 cm$^2$ V$^{-1}$s$^{-1}$.  Supplementary Figure~\ref{fig:mobility}(d) shows a plot of mobility versus charge carrier density ($n$), the data was plotted using the equation $\mu=\sigma/ne$, where $n=C(V_\mathrm{bg}-V_\mathrm{d})/e$ and $\sigma$ is conductivity. One can see that away from the Dirac point, the mobility is nearly independent of $n$.

\begin{figure}[t]
	\begin{center}
		\includegraphics[width=0.7\textwidth]{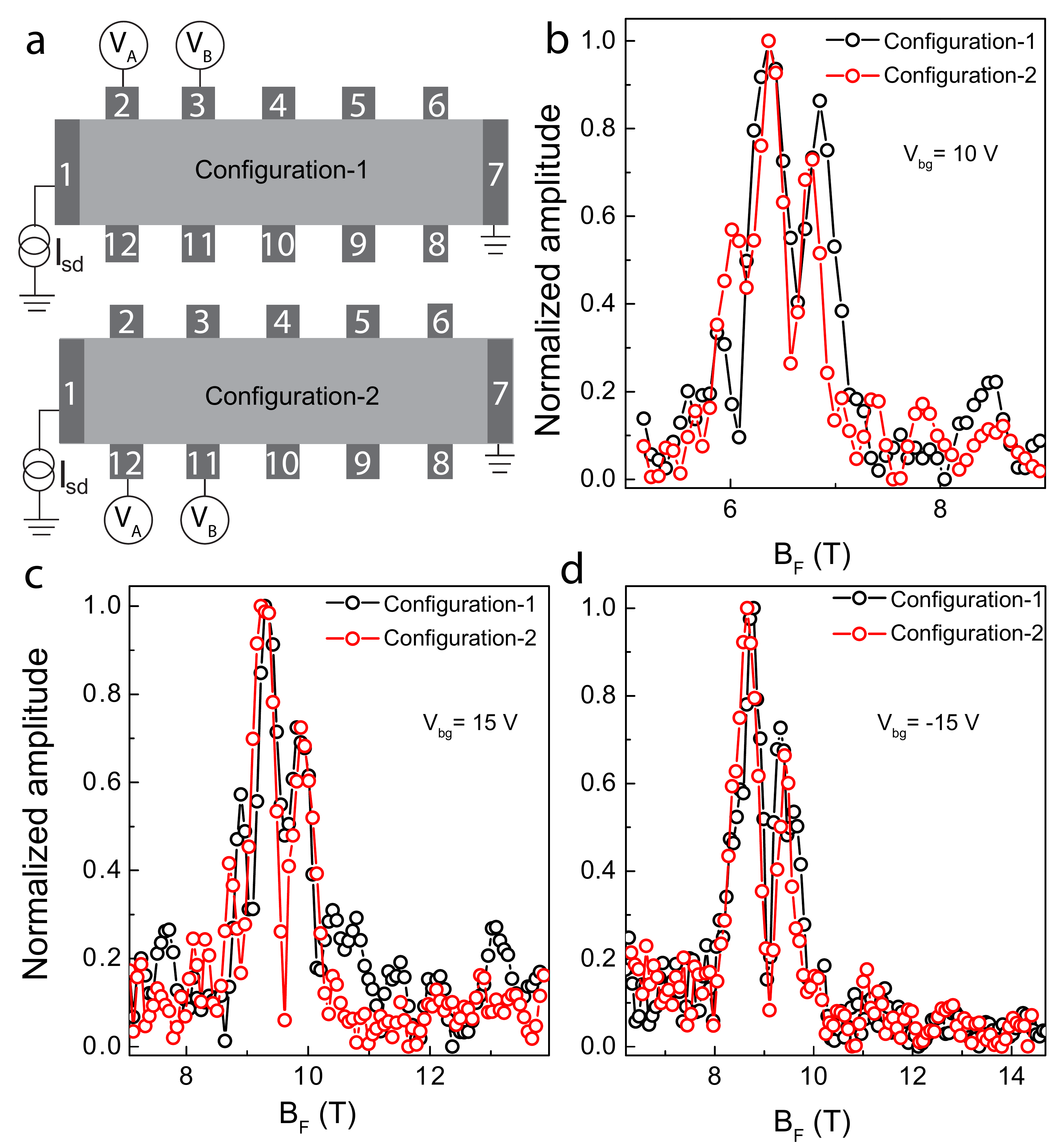}
		\small{\caption{\textbf{FFT of SdH oscillations measured in different configurations.} (a) Measurement configurations for the  data shown in (b), (c) and (d). (b) The FFT spectra for two distinct configurations shown in (a) for $V_\mathrm{bg}=10$~V, (c) for $V_\mathrm{bg}=15$~V, and (d) for $V_\mathrm{bg}=-15$~V.}
			\label{fig:SdH_FFT_Diff_confi}}
	\end{center}
\end{figure}
\begin{figure}[t]
	\begin{center}
		\includegraphics[width=0.65\textwidth]{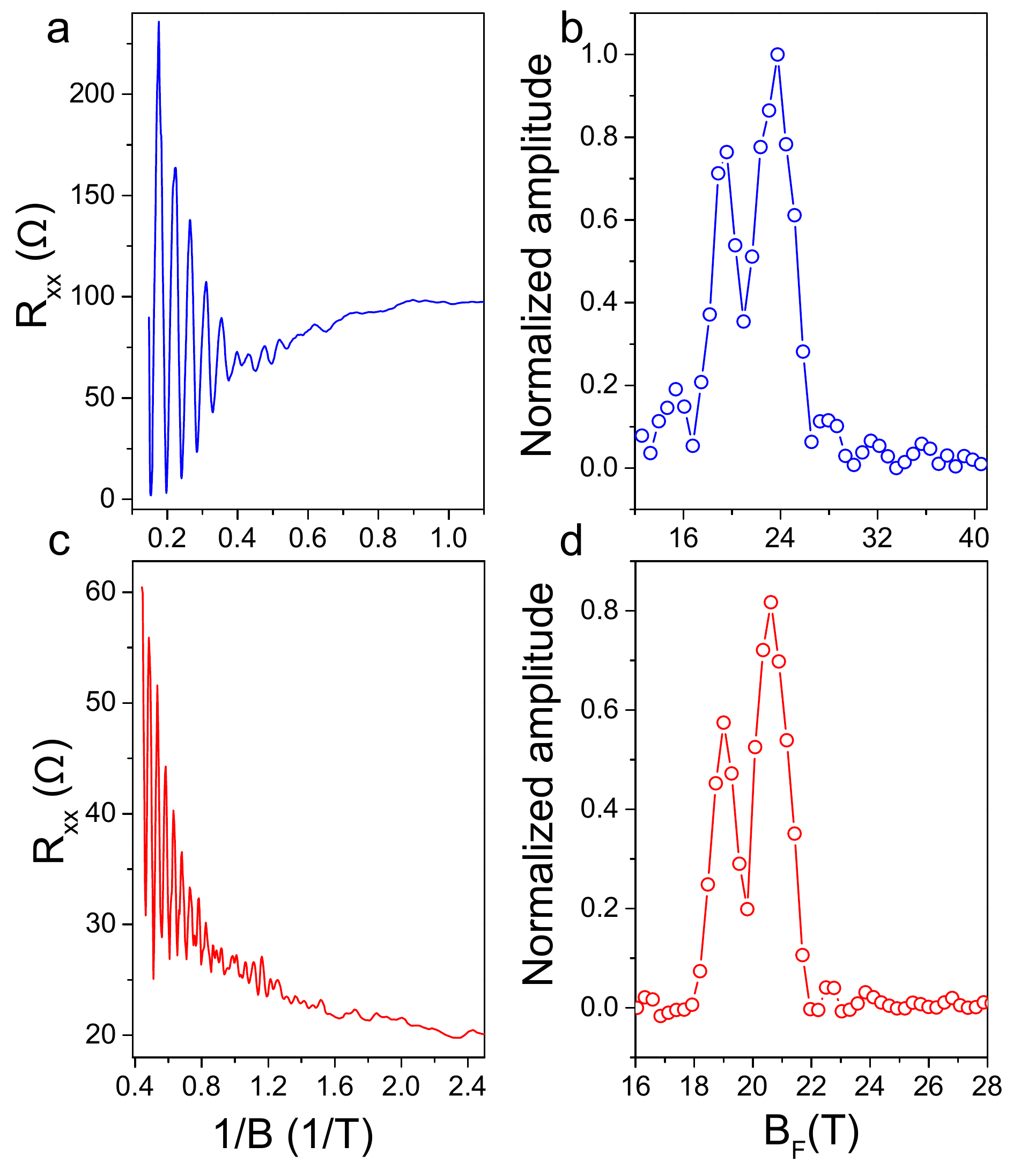}
		\small{\caption{\textbf{SdH oscillations in device B6S1} (a) Plot of the Shubnikov-de Haas oscillations measured on device B6S1 (graphite gated) at $V_\mathrm{bg} = 7$~V and $T = 20$~mK. (b) The corresponding FFT spectra showing two distinct peaks. (c) Plot of the Shubnikov-de Haas oscillations measured on device B6S3 at $V_\mathrm{bg} = -30$~V and $T = 20$~mK. (d) The corresponding FFT spectra showing two distinct peaks.}
			\label{fig:Data_B6S1}}
	\end{center}
\end{figure}

\section{Data for additional measurement configurations and additional devices}

{We measured the Shubnikov-de Haas oscillations (SdH) in multiple configurations in device B9S6. The frequencies extracted for  different configurations via FFT match very well with experimental error bars. The data for two of the configurations are shown in Supplementary Figure~\ref{fig:SdH_FFT_Diff_confi}. In Supplementary Figure~\ref{fig:SdH_FFT_Diff_confi} (a), the measurement configurations are shown for the data in (b), (c) and (d). The FFT for the SdH oscillation measured at $V_\mathrm{bg}=10,15,-15$~V in the two configurations are shown in (b), (c), and (d) respectively. One can note that the FFT for different configurations give the same result. This establishes that the observed two frequencies in the oscillation are intrinsic to the device and do not originate from spatial inhomogeneity of charge carrier density.}

{We measured SdH in several devices, the results from all of them were similar. Here, we present the data from two such devices: (1) Device B6S1, which is heterostructure of SLG and single layer \ch{WSe2} encapsulated in hBN with a graphite back gated and (2) device B6S3, which is an SLG/few-layer-\ch{WSe2} heterostructure encapsulated by hBN, this device is on \ch{SiO2}.   Supplementary Figure~\ref{fig:Data_B6S1}(a) shows the plot of four-probe resistance as function of $1/\mathrm{B}$ at $V_\mathrm{bg}=7$~V and $T=$ 20~mK for the device B6S1, Supplementary Figure~\ref{fig:Data_B6S1}(b) shows the  FFT of the data. The data for device B6S3 measured  at $V_\mathrm{bg}=-30$~V are shown in Supplementary Figure~\ref{fig:Data_B6S1}(c) and (d) respectively. For both devices we observe clear beating of the SdH oscillations and two frequencies in the FFT.} 

\section{Calculation of the dispersion relation}

We extract the effective mass $m^*$ by fitting the normalized amplitude of longitudinal resistivity to the relation~\cite{kuppersbusch2017modifications,lifshitz1956theory}:
\begin{equation}
	\frac{\Delta R_\mathrm{xx}}{R_\mathrm{0}}\propto\frac{2\mathrm{\pi}^2k_\mathrm{B}Tm^{\ast}/\hbar \mathrm{e}B}{sinh\left(2\mathrm{\pi}^2k_\mathrm{B}Tm^{\ast}/\hbar \mathrm{e}B\right)}, 
	\label{eqn_Rxx_ampli}
\end{equation}
where $\Delta R_\mathrm{xx}$ is the amplitude of longitudinal resistivity and $R_\mathrm{0}$ longitudinal resistivity at zero magnetic fields.\\
The effective mass can be written as
\begin{eqnarray}
	&{m^*}^{-1}=&\frac{1}{{\ \hbar}^2}\frac{\partial^2E}{\partial k^2}, \nonumber \\ 
	&{m^\ast}^{-1}=&\frac{1}{{\ \hbar}^2}\frac{\partial}{\partial k}\left(\frac{\partial E}{\partial A_\mathrm{k}}\times\frac{\partial A_\mathrm{k}}{\partial k}\right).
\end{eqnarray}      
Using $A_\mathrm{k}=\mathrm{\pi} k^2$, $2\mathrm{\pi} kdk=\partial A_\mathrm{k}$ yields
\begin{eqnarray}
	&{m^*}^{-1}=&\frac{2\mathrm{\pi}}{{\ \hbar}^2}\frac{\partial E}{\partial A_\mathrm{k}}, \nonumber \\
	&m^*=&\frac{{\ \hbar}^2}{2\mathrm{\pi}}\frac{\partial A_\mathrm{k}}{\partial E}.
\end{eqnarray}
For $A_\mathrm{k}=\mathrm{\pi} k^2$ and for a linear dispersion of single layer  graphene ($E=\hbar v_\mathrm{F} k$),
we have
\begin{eqnarray}
	&A_\mathrm{k}=&\mathrm{\pi}\left(\frac{E}{\hbar v_\mathrm{F}}\right)^2, \nonumber \\
	&m^\ast=&\frac{E}{v_\mathrm{F}^2}={\ \hbar}\frac{k}{v_\mathrm{F}}, \nonumber \\
	&m^\ast=&{\ \hbar}\frac{\sqrt{\mathrm{\pi} n}}{v_\mathrm{F}}=\frac{{\sqrt{\mathrm{\pi}}\ \hbar}}{v_\mathrm{F}}\sqrt n .
\end{eqnarray}
One can fit the experimentally obtained dependence of $m^*$ on $n$ using the relation $m^*=\sqrt{\mathrm{\pi}}\hbar n^\alpha/v_\mathrm{F}$~\cite{neto2009electronic} keeping $\alpha$ and $v_\mathrm{F}$ as fitting parameters. The fits to the experimental data shown in main text (Fig.~\ref{fig:figure3}(c)) yield  $\alpha=0.5\pm 0.02$ and $v_\mathrm{F} = 1.29 \pm 0.04 \times {10}^{6}$~ms$^{-1}$. The value of $\alpha$ is nearly $0.5$, establishing the dispersion relation between energy and momentum for SLG/\ch{WSe2} to be linear~\cite{novoselov2005two}.

\section{Theoretical Modelling}
\begin{figure}[t]
	\begin{center}
		\includegraphics[width=1\textwidth]{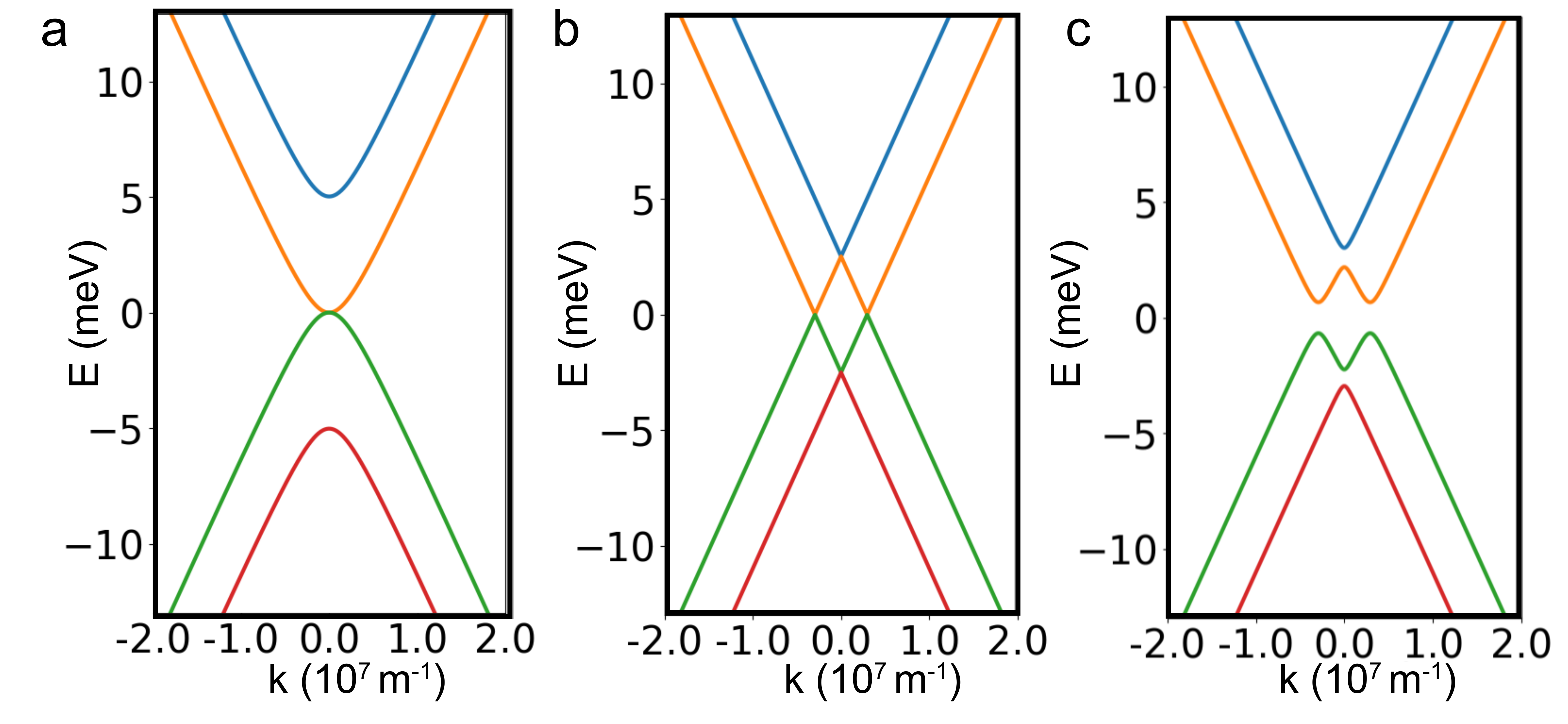}
		\small{\caption{ \textbf{Dispersion relation for different values of $\theta$, $\lambda_\mathrm{R}$ and $\lambda_\mathrm{VZ}$} (a) Plot of the dispersion relation for $\theta=\mathrm{\pi}/2$ ($\lambda_\mathrm{R}= 2.51$ meV and $\lambda_\mathrm{VZ}= 0$ meV). (b) Plot of the dispersion relation for $\theta=0$ ($\lambda_\mathrm{VZ}= ~2.51$ meV and $\lambda_\mathrm{R}= 0$ meV). (c) Plot of the dispersion relation for values of the SOC terms reported in literature: $\lambda_\mathrm{R}= 0.56$ meV and $\lambda_\mathrm{VZ}=2.45$ meV ($\theta =13^\circ)$. }
			\label{fig:sup_fig1}}
	\end{center}
\end{figure}
\begin{figure}[h]
	\begin{center}
		\includegraphics[width=0.8\textwidth]{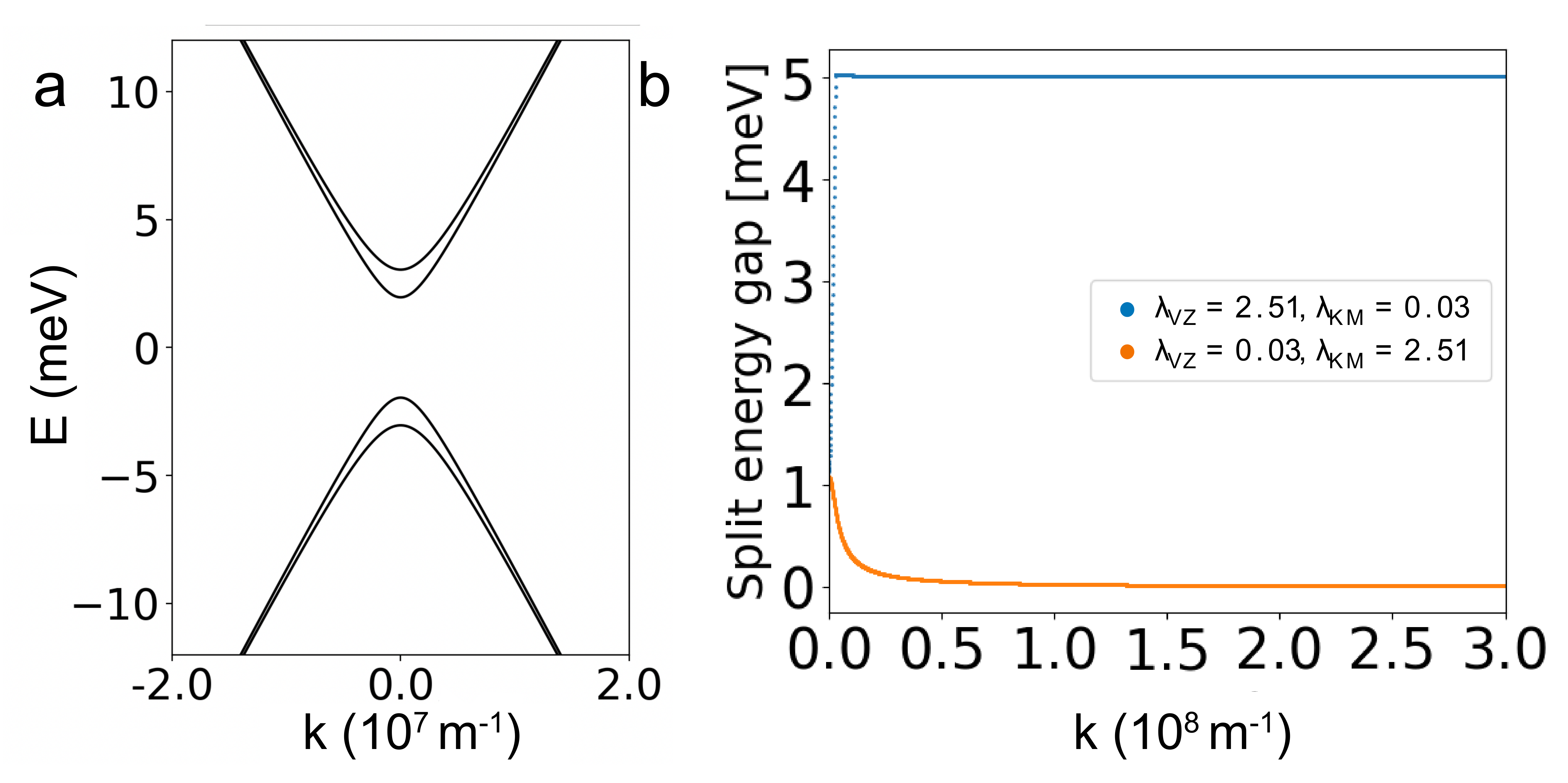}
		\small{\caption{\textbf{Calculated dispersion relation} (a) Plot of the dispersion relation for $\lambda_\mathrm{KM}= 2.51$ meV and $ \lambda_\mathrm{VZ}= 0.03$ meV. As one can see, this does not produce a significant energy gap between the spin-split bands.  (b) Energy gaps between the spin-split bands calculated with the values of $\lambda_\mathrm{VZ}$ and $\lambda_\mathrm{KM}$ interchanged. The blue curve  is calculated for $\lambda_\mathrm{VZ}= 2.51$~meV and $\lambda_\mathrm{KM} = 0.03$~meV whereas the orange curve is calculated for  $\lambda_\mathrm{VZ}= 0.03$ meV and $\lambda_\mathrm{KM} = 2.51$~meV.}
			\label{fig:sup_fig2}}
	\end{center}
\end{figure}

%\begin{figure}[h]
%	\begin{center}
	%		\includegraphics[width=0.7\textwidth]{suppl_3.pdf}
	%		\small{\caption{{The increase in spin- split energy gap as one approaches Dirac point (as observed in our measurements), can be modelled with large values of $\lambda_{KM}$ and $\Delta$. In this plot, we have $\lambda_{KM}$ and $\Delta$ $=20$ meV.}}
		%			\label{fig:sup_fig3}}
	%	\end{center}
%\end{figure}

The continuum Hamiltonian near the Dirac points used for fitting the experimental data has the following terms,
\begin{eqnarray}
	H& =& \hbar v_\mathrm{F} (\eta k_\mathrm{x} \sigma_\mathrm{x} + k_\mathrm{y} \sigma_\mathrm{y})+ \Delta \sigma_\mathrm{z} + \lambda_\mathrm{KM} \eta S_\mathrm{z} \sigma_\mathrm{z}  + \lambda_\mathrm{VZ} \eta S_\mathrm{z} +  \lambda_\mathrm{R} (\eta S_\mathrm{y} \sigma_\mathrm{x} - S_\mathrm{x}\sigma_\mathrm{y}) \nonumber \\ &&+ \frac{\sqrt{3}a}{2} [\lambda_\mathrm{PIA}^{A} (\sigma_\mathrm{z} +\sigma_\mathrm{0}) + \lambda_\mathrm{PIA}^{B}(\sigma_\mathrm{z} -\sigma_\mathrm{0})] (k_\mathrm{x} S_\mathrm{y} - k_\mathrm{y} S_\mathrm{x}). \label{equation9}
\end{eqnarray}
The best fit gives hopping parameter $t= 3979.10\pm 3.99$ meV implying a large Fermi velocity $v_\mathrm{F} = 3ta/2$ in this sample. 
We further note that the parameters in the Hamiltonian which give the spin-split band gap in both conduction and valence bands are  $\lambda_\mathrm{VZ}$ and $\lambda_\mathrm{R}$. The other parameters do not significantly alter the dispersion in the region of experimental data. 

We also find that the best fit gives the values of $\lambda_\mathrm{VZ}$ and $\lambda_\mathrm{R}$ to lie on a circle of radius $2.51$ meV giving a spin-split band gap of $5$ meV, such that 
\begin{equation} \label{equation10}
	\lambda_\mathrm{VZ}= 2.51 ~\cos\theta ~~\rm{meV}~~~ {\rm and}~~~ \lambda_\mathrm{R} = 2.51 ~\sin\theta ~~\rm{meV}.
\end{equation}
The two extreme cases are given by $\theta= 0$ with only valley-Zeeman SOC and $\theta= \mathrm{\pi}/2$ with only Rashba SOC are shown in Supplementary Figure ~\ref{fig:sup_fig1}. 
In absence of experimental data to fix $\theta$, we take the value of $\lambda_\mathrm{R}$ from literature and produce a plot for $\lambda_\mathrm{R}= 0.56$ meV and $\lambda_\mathrm{VZ}=2.45$ meV as shown in Supplementary Figure~\ref{fig:sup_fig1} (c). The actual dispersion of this sample will have a form very similar to  Supplementary Figure~\ref{fig:sup_fig1} (c).

Coming to the effect of the Kane-Mele term on the spin-splitting, Supplementary Figure~\ref{fig:sup_fig2}(a) shows that $\lambda_\mathrm{KM}$ produces a very small energy gap between the spin-split bands, that too only at low-values of momenta $\bold{k}$. It cannot reproduce the band splitting of $\sim 5$ meV obtained from the experimental data. The experimentally obtained spin-splitting energy gap of $\sim 5$ meV  can only be obtained by turning on the Valley-Zeeman and/or the Rashba SOC. This is demonstrated in Supplementary Figure~\ref{fig:sup_fig2}(b) by interchanging the values of $\lambda_\mathrm{VZ}$ and $\lambda_\mathrm{KM}$. One can see that while $\lambda_\mathrm{VZ}=2.51$~meV and $\lambda_\mathrm{KM} =0.03$ produces a spin-splitting of about 5~meV, for $\lambda_\mathrm{VZ}=0.03$ and $\lambda_\mathrm{KM} =2.51$~meV the splittng is negligibly small away from the Dirac point. This establishes that the energy gap between the  spin-split bands is determined by the strength of $\lambda_\mathrm{VZ}$ and $\lambda_\mathrm{R}$, as discussed in the main text. 

Interestingly, it is observed from the experimental data that as one approaches Dirac point, the energy gap between the spin-split bands increases. To reproduce this pattern in theoretical model,  we need to use significantly large values of $\lambda_\mathrm{KM}$ and $\Delta$  $(\approx  20 meV )$ along with the Valley-Zeeman and Rashba SOC. Such large values of $\lambda_\mathrm{KM}$ and $\Delta$ are not reasonable and have not been reported so far; $\lambda_\mathrm{KM} \approx 0.03$ meV from most reports~\cite{garcia2018spin,gmitra2016trivial}. Thus, although it is tempting to explain the trend of increase in the energy gap between the spin-split bands as one approaches the Dirac point to originate from a large $\lambda_\mathrm{KM}$, we refrain from doing so.

%	\bibliography{arxiv}

\end{document}